\documentstyle[12pt]{article}
\newcommand{\half}{\frac 1 2 }

\newcommand{\be}{\begin{eqnarray}}
\newcommand{\ee}{\end{eqnarray}}

\catcode`\@=11
\newcommand{\ccite}[1] {\@ifundefined{b@#1}{\bf ?}{\@nameuse{b@#1}}}
\catcode`\@=12

\begin{document}
\vspace*{1cm}
\centerline{\Large\bf Algebra of one-particle operators for the Calogero
model}
\vspace* {-45 mm}
\begin{flushright} Oslo SHS-95-1 \\
 October 1995 \\
hep-th/9510184
\end{flushright}
\vskip 55mm
\centerline{\bf Serguei B. Isakov$^{\star}$ and Jon Magne Leinaas$^{\dagger}$}
\medskip
\centerline{Centre for Advanced Study}
\centerline{at the Norwegian Academy of Science and Letters,}
\centerline{P.O. Box 7606 Skillebekk, N-0205 Oslo, Norway} \vskip 15mm
\centerline{\bf ABSTRACT}
 \medskip
An algebra ${\cal G}$ of symmetric {\em one-particle}
operators is constructed for the Calogero model. This is an
infinite-dimensional
Lie-algebra, which is independent of the interaction parameter $\lambda$ of
the model.
It is constructed in terms of symmetric polynomials of raising and lowering
operators
which satisfy the commutation relations of the $S_N$-{\em extended}
Heisenberg algebra.
We interpret ${\cal G}$ as the algebra of observables for a system of
identical particles
on a line. The parameter $\lambda$, which characterizes (a class of)
irreducible
representations of the algebra, is interpreted as a statistics parameter for
the
identical particles. \vskip 3mm

\vfil
\noindent
$^{\star}$On leave from the Medical Radiology Research Center, Obninsk,
Kaluga Region 249020, Russia. E-mail: ISAKOV@SHS.NO\\
$^{\dagger}$Permanent address: Institute of Physics, University of Oslo, PO
Box 1048 Blindern, N-0316 Oslo, Norway. E-mail: J.M.LEINAAS@FYS.UIO.NO\\

\eject
\section{Introduction}

This work is motivated, on one hand, by the algebraic approach to identical
particles
(Heisenberg quantization) \cite{LM-Heis}, on the other hand, by the recent
progress in
understanding of the algebraic properties of the Calogero model
in terms of the $S_N$-{\em extended} Heisenberg algebra
\cite{V,Poly-PRL92,BHV}. In
its general form the main idea of the algebraic approach can be formulated
as searching
for an algebra (the {\em algebra of observables} for a system of identical
particles)
which should have the same form independent of the statistics of particles. The
statistics of particles is assumed to arise at the level of irreducible
representations
of the algebra of observables: different irreducible representations
correspond to
different statistics. Structures possessing these properties were found for two
particles in one and two spatial dimensions.

For two particles on a line, the algebra of observables is the algebra
$sl(2,{\bf R})$
\cite{LM-Heis}. The irreducible representations of the same algebra also
classify the
solutions of the Calogero model for two particles \cite{Perelomov}, and the
algebraic
approach suggests the interpretation of the singular $1/x^2$-potential of
the Calogero
model as a ``statistical'' interaction between the particles, which
introduces fractional
statistics in one dimension \cite{LM-Heis} (see also discussion in terms of
the Schr{\"o}dinger quantization in Ref.~\cite{PolyNPB}).
Note that the equivalence of models with $1/x^2$ interaction
to systems of non-interacting particles obeying
so-called exclusion statistics \cite{H} has also been discussed
recently \cite{StatMech}.

The algebra $sl(2,{\bf R})$ can be constructed in terms of one-particle
operators
generated by the $S_N$-extended Heisenberg algebra of the Calogero model
\cite{BHV}. In this paper we use this extended Heisenberg algebra to
construct an
algebra of one-particle operators for an arbitrary number of particles in
the Calogero
model. The algebra is independent of the interaction parameter ({\it
statistics}
parameter) of the Calogero model, and the parameter then characterizes
different
representations of the algebra. In this way our paper can be viewed as a
realization of the
algebraic approach to fractional statistics for spinless identical
particles on a line.
Apart from this interpretation the presence of this algebra should also be
of interest
for the study of the algebraic structure of the Calogero model.

We should add to this that we are not able to give an expression for the
parameter\-independent algebra in a closed form. We rather show the
existence of this
algebra and give a systematic way to generate a basis for the algebra. This is
supplemented by a detailed discussion of how this is done in some specific
cases.

\section{Definition of the algebra}

We start with a system of $N$ classical particles, with the coordinates
$\{x_i\}$
($i=1,2,...,N$) on a line. Let $\{p_i\}$ be the set of canonical momenta.
Observables
for a system of non-interacting identical particles should be symmetric
under particle
permutations and can be generated {\it e.g.} by the symmetric polynomials
of the form
$s_k=\sum_{i=1}^N \xi_i^k$, where $\xi_i=(x_i,p_i)$ are coordinates in the
phase space
\cite{LM-Heis}. Also in the quantum case the observables should be
symmetric, and they
can be generated by the same symmetric polynomials, but now with $x_i$ and
$p_i$ as
operators of the $N$-particle system.

The operator ordering problem prevents a unique mapping between the
classical and
quantum cases. The algebraic relations between the observables are
different in the two
cases. In the classical case the symmetric polynomials define, by Poisson
brackets, an
infinite-dimensional Lie algebra referred to as $w_\infty$. In the quantum
case the
symmetric polynomials define a commutation algebra referred to as
$W_{1+\infty}$ (see
{\em e.g.} \cite{CTZ}). The latter can be viewed as containing quantum
corrections (higher order in $\hbar$) relative to the classical algebra
$w_\infty$.
$W_{1+\infty}$ is an algebra of observables for fermions as well as
bosons. However, for
the case of generalized statistics which we examine here, this algebra is
not represented.
Instead we find a larger algebra of observables, which is homomorphic to
$w_\infty$ as well as to $W_{1+\infty}$.

To construct this algebra, we use the $S_N$-{\it extended} Heisenberg
algebra which is
generated by the operators $a_i$, $a_i^\dagger$, and $K_{ij}$ satisfying
the relations
\cite{V,Poly-PRL92,BHV}
\be
[a_i, a_j]=[a_i^{\dagger}, a_j^{\dagger}]=0\,, \;\; [a_i,
a_j^{\dagger}]=\delta_{ij}(1+\lambda \sum_l K_{il}) -\lambda K_{ij}\;,
\label{1}\ee
\be
K_{ij}K_{jl}=K_{jl}K_{il}=K_{il}K_{ij} \;\;\;\;{\rm for} \;\;\; i\neq
j,\; i\neq l, \;j\neq l\,,
\label{3}\ee
\be
K_{ij}K_{mn}=K_{mn}K_{ij} \;\;\;\; {\rm for} \;\; i, j, m, n \;\;{\rm all \;
different}\,,
\label{4}\ee
\be
K_{ij}a_j=a_iK_{ij}\;,\;\; K_{ij}a_j^{\dagger}=a_i^{\dagger}K_{ij}\,.
\label{5}\ee
The operators $K_{ij}$, which generate a representation of the symmetric group
$S_N$, are defined for
$i\neq j$, with
$K_{ij}=K_{ji}$ and
$(K_{ij})^2=1$. For convenience, we may define $K_{ii}=0$. In terms of the
above
operators the Hamiltonian of the {\em extended} Calogero system is defined as
\be
H=\frac12 \sum_{i=1}^N\{a_i,a_i^\dagger\}\,.
\label{ham}\ee
This operator has the form of
a generalized system of harmonic oscillators, and the eigenvalue problem of
$H$ is
easily solved, with the operators $a_i^\dagger$ as raising operators in the
spectrum \cite{BHV}.

The original Calogero problem \cite{Cal}, which describes $N$ identical
particles on the line, with a harmonic oscillator potential
and a (``statistical") pair interaction
$V(x_i-x_j)=\lambda (\lambda -1)/(x_i-x_j)^2$, corresponds to the subspace
of solutions
which are totally symmetric (or totally antisymmetric) with respect to particle
permutations. (To be precise, this is the form of the wave functions after
a singular
factor of the form $\prod_{i>j}(x_i-x_j)^{\lambda}$ has been factored out
of the wave
functions.) The symmetric functions of the operators $a_i$ and
$a_i^\dagger$ are the
observables of this system.

For the discussion below it is convenient to introduce the
following notation. We define a {\it single-particle} operator ${\cal A}_i$
as a linear
combination of products of operators $a_i$ and $a_i^{\dagger}$ with the
same index,
e.g. ${\cal A}_i=2a_ia_i^{+2}a_i^3+a_i^{+4}a_i^2$. {\it One-particle}
operators are
obtained from single-particle operators by summation over all particles,
{\em i.e.} they have
the form $\sum_i {\cal A}_i$ \cite{LM-Heis}. The one-particle operators are
observables
of this system of identical particles, and the algebra we seek is an
algebra generated
by these operators, which is independent of the statistics parameter
$\lambda$. This
algebra generalizes the algebra $W_{1+\infty}$ of bosons and fermions.

To construct the algebra, we first verify the following important identity
\be
\sum_{ij} {\cal A}_j [a_i^r,a_j^{\dagger}]{\cal A}'_j=\sum_i {\cal A}_i
ra_i^{r-1}{\cal A}'_i \,,
\label{7}\ee
where $ {\cal A}_i$ and ${\cal A}'_i$ are arbitrary single-particle
operators.
Starting from the last relation in (\ref{1}), we derive inductively that
\be
[a_i^r,
a_j^{\dagger}]=r \delta_{ij} a_i^{r-1} + \lambda
\delta_{ij}\sum_l\sum_{s=0}^{r-1}a_i^{r-1-s}a_l^s K_{il} -\lambda
\sum_{s=0}^{r-1}a_i^{r-1-s}a_j^s K_{ij} \,.
\label{6} \ee
Inserting (\ref{6}) into the left-hand side of (\ref{7}), we get
\begin{equation}
\sum_{ij} {\cal A}_j [a_i^r,a_j^{\dagger}]{\cal A}'_j=\sum_i {\cal A}_i
ra_i^{r-1}
{\cal A}'_i +\lambda \sum_{il}{\cal A}_i\sum_{s=0}^{r-1}a_i^{r-1-s}a_l^s
K_{il}{\cal A}'_i -\lambda \sum_{ij}{\cal A}_j\sum_{s=0}^{r-1}a_i^{r-1-s}a_j^s
K_{ij}{\cal A}'_j
\label{8}
\end{equation}
The terms containing the operators $K_{ij}$ are canceled, and
the identity (\ref{7}) follows. This identity is central for the construction
of
the statistics-independent algebra.

We now consider the operators of the form $L_{0n}=\sum_i a_i^n$ and
$L_{m0}=\sum_i
a_i^{{\dagger}m}$ . For the commutator we find \be
[L_{0n},L_{m0}]=\sum_{ij} [a_i^n,a_j^{{\dagger}m}]=
\sum_{ij}\sum_{s=0}^{m-1}a_j^{{\dagger}s}[a_i^n,a_j^{\dagger}](a_j^{\dagger}
)^{m-s-1}=
\sum_{ij}\sum_{r=0}^{n-1}a_i^r[a_i,a_j^{{\dagger}m}]a_i^{n-r-1}\,.
\label{9}\ee Using then the
identity (\ref{7}), we obtain
\be[L_{0n},L_{m0}]=n\sum_i\sum_{s=0}^{m-1}a_i^{{\dagger}s}a_i^{n-1}
(a_i^{\dagger})^{m-s-1}=
m\sum_i\sum_{r=0}^{n-1}a_i^r(a_i^{\dagger})^{m-1} a_i^{n-r-1}\,.
\label{10}\ee
which
shows that the commutator is a one-particle operator. For the special case
$n=2$, we
have
\be [L_{02},L_{m0}]=2mL_{m-1,1}\,,
\label{11}\ee
where $L_{m1}$ is defined by
\be
L_{m1}=\half\sum_i (a_i^{{\dagger}m} a_i + a_i a_i^{{\dagger}m}) \,.
\label{12}\ee

It is now straightforward to show that the operators of the form $L_{m0}$
and $L_{m1}$ form a closed algebra. We first note that
\be [L_{m0},L_{n0}]=0 \,,
\label{13}\ee due to
the commutativity of the operators $a_i^{\dagger}$ with arbitrary indices.
Next, using the identity (\ref{7}), we get
\be [L_{m1},L_{n0}]=nL_{m+n-1,0}\,.
\label{14} \ee
Finally, for calculating the commutator $[L_{m1},L_{n1}]$, we rearrange
it, using the Jacobi identity, as follows:
\be
2(m+1)[L_{m1},L_{n1}]=[[L_{02},L_{m+1,0}],L_{n1}]=
[L_{02},[L_{m+1,0},L_{n1}]]-[L_{m+1,0},[L_{02},L_{n1}]]\,.
\label{14a}\ee
Then,  using
the identity (\ref{7}) and its conjugate, and, in addition, taking into
account the relation
\be
\sum_{i}\left( a_i^{\dagger k}a_ia_i^{\dagger k'}
+a_i^{\dagger k'}a_ia_i^{\dagger k}\right)=
\sum_{i}\left(a_ia_i^{\dagger k+k'}+a_i^{\dagger k+k'}a_i\right) \,,
\label{15}
\ee
for $k$ and $k'$ non-negative integers, which is derived from (\ref{6})
straightforwardly, we obtain
\be
[L_{m1},L_{n1}]=(n-m)L_{m+n-1,1} \,.
\label{16}
\ee
Note that Eqs.~(\ref{13}), (\ref{14}), and (\ref{16}) are the commutation
relations for (the
positive-frequency part of) a $U(1)$ current algebra. The Hermitian
conjugation yields
one more $U(1)$ current algebra with operators $L_{0m}\equiv L_{m0}^\dagger$
and
$L_{1m}\equiv L_{m1}^\dagger$. The operators $L_{m0}$ (or $L_{0m}$) define
a $U(1)$
Kac-Moody (sub)algebra, and the operators $L_{n1}$ (or $L_{1n}$) define a
Virasoro
(sub)algebra. For brevity we shall refer to the operators $L_{m0}$ and
$L_{0m}$ as the
Kac-Moody (KM) operators and the operators $L_{m1}$ and $L_{1m}$ as the
Virasoro
operators. Note that the Virasoro algebra generated from the $S_N$-extended
Heisenberg
algebra was also found by Bergshoeff and Vasiliev \cite{BV} and by
Polychronakos
\cite{Poly-pc}.

The procedure of derivation of the commutation relations between the
operators $L_{m0}$
and $L_{n1}$ is part of the general construction. We seek an algebra with
generators of
the form $L_{mn}$, where $m$ and $n$, which refer to powers of the operators
$a_i^{\dagger}$ and $a_i$, now can take arbitrary non-negative values. The
set of
operators is generated from the KM operators by repeated commutators. One
building
block of this construction is the observation that the commutator of a KM
operator with
any one-particle operator is another one-particle operator,
\be
[L_{0n},\sum_{i}{\cal
A}_i]= \sum_{i}{\cal A'}_i
\ee
(with a similar expression valid for $L_{m0}$). This
follows directly from the identity (\ref{7}). Indeed, the commutator between
$L_{0n}=\sum_j a_j^n$ and $\sum_i{\cal A}_i$ can be written as a sum of
terms where
$a_j^n$ is commuted with each factor
$a_i^{\dagger}$ in the operator $\sum_{i}{\cal A'}_i$. For each of these
terms the
identity (\ref{7}) can be applied. The result is a sum of one-particle
operators, which
is again a one-particle operator.

To proceed, we consider arbitrary `strings' of consecutive commutators, of
the form
$[L_N\cdots ,[L_3,[L_2,L_1]]\cdots]$, where $L_1,L_2,\dots$ are KM
operators. From the
discussion above it follows that such a string is a one-particle operator.
Furthermore,
the set of such strings is closed under commutation, {\em i.e.} the
commutator of two
strings is a linear combination of strings. This can be shown by re-writing the
commutator with the use of the Jacobi identity. The infinite-dimensional
Lie algebra
generated from the KM operators in this way, which will be denoted by
${\cal G}$, is
the one we seek.

One more definition will be useful. We call an operator {\it
mirror-symmetric} to a
given product of the operators $a_i$ and $a_i^{\dagger}$ if it is obtained
by reversing
the order of operators $a_i$ and $a_i^{\dagger}$ in the product. This
definition is
naturally extended to one-particle operators. For example, the operator $\sum_i
a_ia_i^{{\dagger}3}a_i^2$ is mirror-symmetric to the operator $\sum_i
a_i^2a_i^{{\dagger}3}a_i$. The commutator of a KM operator with a
mirror-symmetric
one-particle operator is another mirror-symmetric operator. This is readily
shown by the
application of identity (\ref{7}). As a consequence of this, all the
operators of the
algebra ${\cal G}$ will be mirror-symmetric. One should note that only a
subset of the
mirror-symmetric operators will be included in the algebra. This will be
shown by
explicit calculation in some specific cases.

Since the algebra ${\cal G}$ is an algebra of one-particle operators, an
arbitrary
element can be written as a linear combination of operators $\sum_i A_i$,
where $A_i$
are products of the the operators $a_j$ and $a_j^{\dagger}$. We say that an
element of
the algebra ${\cal G}$ is of {\it order} $(m,n)$ if all $A_i$ are products
of $m$
operators $a_i^{\dagger}$ and $n$ operators $a_i$, and we write such an element
generically as $L_{mn}$. If the commutators are evaluated by use of the
identity
(\ref{7}), as outlined above, a string of KM operators will give rise to a
one-particle
operator of order $(m,n)$, where
\be
m= \sum_i m_i-N+1\,, \nonumber \\
n=\sum_i n_i-N+1 \,,
\ee
with $(m_i,0)$ and $(0,n_i)$ as the orders of the KM operators and $N$
as the
number of operators in the string. (A negative value of $m$ or $n$ then
corresponds to
a vanishing operator.) It follows that operators $L_{mn}$ of all possible
orders
$(m,n)$ with
$m\geq 0$ and
$n\geq 0$ will be generated in this way. However, several operators of the
same order
$(m,n)$, but differing in the ordering of the operators $a_i^{\dagger}$ and
$a_i$ may
be generated. To define the algebra more precisely, we have to specify the
relation
between these operators.

We have already noted an important identity (\ref{15}) between operators
with different
orderings of the operators $a_i^{\dagger}$ and $a_i$. This identity reduces
the number
of independent operators of order $(m,1)$ (or $(1,n)$) to one.
{}From this identity other identities can be deduced by commutation with KM
operators and
by the use of (\ref{7}). We note in particular the useful identity
\be
\sum_i [{\cal A}_i,[a_i,a_i^{\dagger}]]=0 \,, \label{15a} \ee This
identity, which is
valid for an arbitrary one-particle operator $\cal A$, can be verified
either by direct
calculation, or, if $\cal A$ belongs to ${\cal G}$, it can be derived from
the identity
(\ref{15}) as discussed above. In fact a more general form of this identity
can be
given, \be \sum_i [{\cal A}_i,[a_i,a_i^{\dagger}]^n]=0 \,, \label{15b} \ee with
$n$ as any positive integer.

When the identities are introduced, the number of linearly independent
operators of
given order $(m,n)$ is reduced. Some of these identities in fact are needed
to satisfy
the Jacobi identity. Others may not follow from consistency of the algebra,
but are
imposed in order to define a {\em minimal} algebra consistent with the
algebraic
relations of the extended Heisenberg algebra. Our precise definition of
${\cal G}$ will
be that all {\it statistics independent} identities between one-particle
operators,
which can be derived from the extended Heisenberg algebra, should be
included in the
definition of the $\cal G$. From studies of the low-order operators it
seems that all
statistics independent identities only relate operators of the same order
$(m,n$), and
in the following we shall assume that this is generally true. Other {\em
statistics
dependent} identities may possibly be derived from the extended Heisenberg
algebra, but
these will not be imposed on the algebra $\cal G$. Instead these identities are
considered as characteristic for certain (irreducible) representations of
the algebra.

We stress the point that the algebra $\cal G$, as defined above, is a
statistics
independent algebra. This follows from the fact that the identity
(\ref{7}), which is
used to evaluate commutators, as well as the other identities which are
used to relate
operators of the same order, do not make any reference to the statistics
parameter
$\lambda$.

Since we cannot give a general list of generators and commutation relations
of the
algebra defined above, our approach will be to look for a systematic way to
generate
operators and commutation relations. The difficult part then will be to
keep track of
the identities, and thus to specify what are the independent elements of
the algebra
for a given order $(m,n)$. One of the complications is that the identities
relate
general one-particle operators which do not necessarily belong to the
algebra. This
means that we have to work in a larger space of operators than is strictly
needed for
the definition of the algebra.

\section{Spin representation}

Before we turn to the question of how to construct the algebra, it is useful
to introduce a classification of the operators of the algebra in terms of
spin. The
operators
$\frac12 L_{02}$, $\frac12 L_{20}$, and$\frac12 L_{11}$ define the
commutation algebra
$su(1,1)$, with $\frac12 L_{02}$, $\frac12 L_{20}$ as lowering and raising
operators,
respectively. After a redefinition, with suitable factors of $i$, they can
also be viewed
as defining the algebra $su(2)$. Consider the action of the operators $L_{02}$,
$L_{20}$, and $L_{11}$ on the elements
$g$ of the algebra ${\cal G}$ in the following way \be L_{02}: \;\; g \to
[L_{02},g]
\label{1b}\ee (and similarly for $L_{20}$ and $L_{11}$). If we consider the
antisymmetric composition of the action of two operators of the algebra
$su(2)$, {\em e.g.}
the operators $L_{02}$ and $L_{20}$, the Jacobi identity gives \be
[L_{02},[L_{20},g]]-[L_{20},[L_{02},g]]=[[L_{02},L_{20}],g] \label{2b} \ee
This shows
that the commutation relations are preserved under mapping of the original
operators
$L_{02}$, $L_{20}$, and $L_{11}$ into operators acting on ${\cal G}$. Thus, the
operators define a representation of the algebra $su(1,1)$ (or $su(2)$) on
the algebra
${\cal G}$. The algebra ${\cal G}$ then can be divided into subspaces
corresponding to
irreducible, and in fact finite dimensional, representations of given spin $s$.

{}From (\ref{6}) we derive
\be
\half \sum_i[a_i^{\dagger}a_i+a_i a_i^{\dagger}, a_j^{\dagger}]=
a_j^{\dagger} \,,
\ee
which implies the following commutator between $L_{11}$ and any operator
$L_{mn}$ of order $(m,n)$,
\be [L_{11},L_{mn}]=(m-n)L_{mn} \,.
\ee
With $\half L_{11}$ interpreted as the z-component of the spin
operator, this gives
the following spin value for an operator of order $(m,n)$
\be
s_z=\frac12 (m-n) \,.
\label{sz}
\ee The operators $L_{20}$ and $L_{02}$ act as raising and lowering operators,
respectively, and by use of these operators, one can construct multiplets
of given
spin. In an $(m,n)$ diagram these multiplets are located along the diagonal
with fixed $m+n$, and lie symmetrically around $m=n$. As a consequence of
this, there
will be one multiplet of maximal spin associated with any point
$(m,n)$ in the diagram. The maximal spin value is
\be
s_{{\rm max}}(m,n)=\half (m+n) \,,
\label{smax}\ee
and this multiplet will include the Kac-Moody as well as Virasoro
operators with the given value of $m+n$.

This structure suggests the interpretation of the degeneracies at given
$(m,n)$ in
terms of different spins. We thus introduce the notation $L_{nm}^{s\alpha}$
with $s$ as
an integer or half-integer smaller or equal to $s_{\rm max}$, given by
(\ref{smax}). The parameter $\alpha$ labels different multiplets with the
same spin at
a given order $(m,n)$. (The parameter $\alpha$ or both parameters $\alpha$
and $s$ may
be dropped in cases where there is no corresponding degeneracy.)

It is worth while noting that there is no multiplet corresponding to $s=s_{{\rm
max}}-1$. This follows from the fact that the there is only one independent
operator of
order $(m,1)$ (or $(1,m)$), and this operator has to belong to the
multiplet $s=s_{{\rm
max}}$. The maximal spin multiplet is unique, but for lower spin more than
one multiplet
with the same spin may appear. From studying the special representation of
${\cal
G}$ characterized by $\lambda=0$ we conclude that there is at least one
multiplet for
each of the spin values $s=s_{\rm max}-2k$, with k as a positive integer
smaller or equal
to $s_{\rm max}/2$ (see Sect. 5). For $m+n \leq 5$ there is no additional
degeneracy, but in the general case there will be more than one multiplet
with the same
spin. The degeneracies derived explicitly in Sect. 4, and listed in Fig.~1,
indicate the
presence of two independent spin
$1$ multiplets for $m+n=6$.


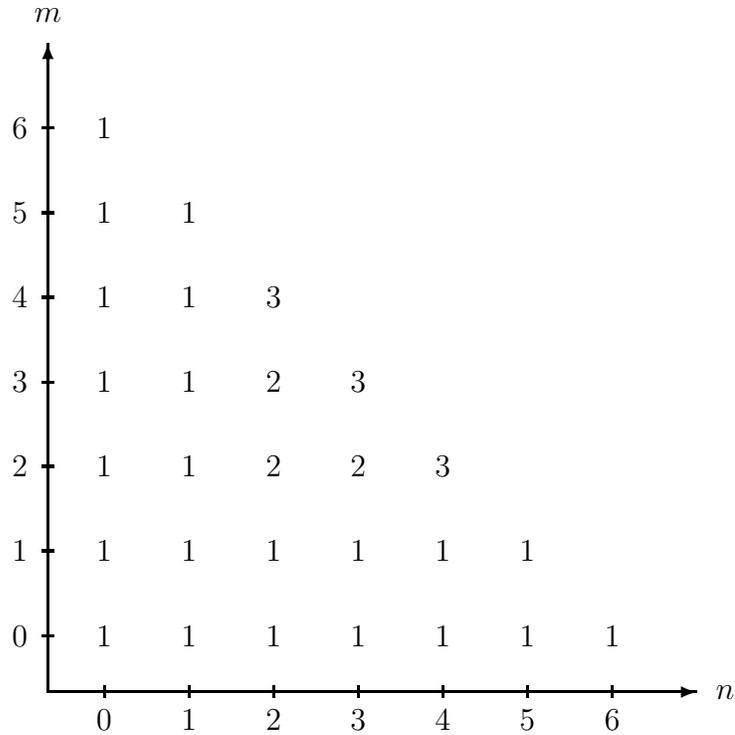
\begin{figure}[htb]
\unitlength=0.75mm
\thicklines
\begin{center}
\begin{picture}(120,130)
\put(10,10){\vector(1,0){115}}
\put(10,10){\vector(0,1){115}}
\put(20,9){\line(0,1){2}}
\put(35,9){\line(0,1){2}}
\put(50,9){\line(0,1){2}}
\put(65,9){\line(0,1){2}}
\put(80,9){\line(0,1){2}}
\put(95,9){\line(0,1){2}}
\put(110,9){\line(0,1){2}}
\put(9,20){\line(1,0){2}}
\put(9,35){\line(1,0){2}}
\put(9,50){\line(1,0){2}}
\put(9,65){\line(1,0){2}}
\put(9,80){\line(1,0){2}}
\put(9,95){\line(1,0){2}}
\put(9,110){\line(1,0){2}}
\put(130,10){\makebox(0,0){$n$}}
\put(10,130){\makebox(0,0){$m$}}
\put(20,20){\begin{picture}(100,100)
\put(0,0){\makebox(0,0){$1$}}
\put(0,-15){\makebox(0,0){$0$}}
\put(15,-15){\makebox(0,0){$1$}}
\put(30,-15){\makebox(0,0){$2$}}
\put(45,-15){\makebox(0,0){$3$}}
\put(60,-15){\makebox(0,0){$4$}}
\put(75,-15){\makebox(0,0){$5$}}
\put(90,-15){\makebox(0,0){$6$}}
\put(-15,0){\makebox(0,0){$0$}}
\put(-15,15){\makebox(0,0){$1$}}
\put(-15,30){\makebox(0,0){$2$}}
\put(-15,45){\makebox(0,0){$3$}}
\put(-15,60){\makebox(0,0){$4$}}
\put(-15,75){\makebox(0,0){$5$}}
\put(-15,90){\makebox(0,0){$6$}}
\put(15,0){\makebox(0,0){$1$}}
\put(30,0){\makebox(0,0){$1$}}
\put(45,0){\makebox(0,0){$1$}}
\put(60,0){\makebox(0,0){$1$}}
\put(75,0){\makebox(0,0){$1$}}
\put(90,0){\makebox(0,0){$1$}}
\put(0,15){\makebox(0,0){$1$}}
\put(15,15){\makebox(0,0){$1$}}
\put(30,15){\makebox(0,0){$1$}}
\put(45,15){\makebox(0,0){$1$}}
\put(60,15){\makebox(0,0){$1$}}
\put(75,15){\makebox(0,0){$1$}}
\put(0,30){\makebox(0,0){$1$}}
\put(15,30){\makebox(0,0){$1$}}
\put(30,30){\makebox(0,0){$2$}}
\put(45,30){\makebox(0,0){$2$}}
\put(60,30){\makebox(0,0){$3$}}
\put(0,45){\makebox(0,0){$1$}}
\put(15,45){\makebox(0,0){$1$}}
\put(30,45){\makebox(0,0){$2$}}
\put(45,45){\makebox(0,0){$3$}}
\put(0,60){\makebox(0,0){$1$}}
\put(15,60){\makebox(0,0){$1$}}
\put(30,60){\makebox(0,0){$3$}}
\put(0,75){\makebox(0,0){$1$}}
\put(15,75){\makebox(0,0){$1$}}
\put(0,90){\makebox(0,0){$1$}}
  \end{picture}}

\end{picture}
\end{center}
\caption{The degeneracies for the  points $(m,n)$ with $m+n\leq 6$.}
\label{fig1}
\end{figure}


The spin representation of the operators implies the following general form
for the
commutators of the algebra $\cal G$,
\be
\left[L^{s\alpha}_{mn},L^{s'\alpha'}_{m'n'}\right]=\sum_{\alpha''}\sum
_{s''=|s-s'|}^{s+s'}d_{s\,s_{\rm max}\,\alpha,s'\,s'_{\rm
max}\,\alpha'}^{s''\alpha''}
\left\langle{{ss's_zs'_z}}
\mathrel{\left | {\vphantom {{ss's_zs'_z} {s''s''_z}}} \right.
\kern-\nulldelimiterspace} {{s''s''_z}}
\right\rangle L^{s''\alpha''}_{m+m'-1,n+n'-1} \,.
\label{5.1} \ee
Here $s_{\rm max}$ and $s_z$
are determined by $m$ and $n$ through eqs. (\ref{smax}) and (\ref{sz}), and
$\left\langle {{ss's_zs'_z}} \mathrel{\left | {\vphantom{{ss's_zs'_z}
{s''s''_z}}}
\right. \kern-\nulldelimiterspace} {{s''s''_z}} \right\rangle$ are
Clebsch-Gordan
coefficients. In addition to the selection rules of the Clebsch-Gordan
coefficients
there exist other selection rules for the coefficients
$d_{s\,s_{\rm max}\,\alpha,s'\,s'_{\rm max}\,\alpha'}^{s''\alpha''}$. One
of these is related
to the fact that there is no multiplet corresponding to $s=s_{\rm max}-1$.
In Appendix B
other cases of vanishing coefficients are mentioned.

In the following we will normalize
the operators within one spin multiplet such that
\be
[L_{02},L_{mn}^{s\alpha}]=
2(s+s_z)L_{m-1,n+1}^{s\alpha} \,.
\label{30}\ee
One should note that this gives a non-standard normalization of the
Clebsch-Gordan coefficients
$\left\langle {{ss's_zs'_z}} \mathrel{\left | {\vphantom{{ss's_zs'_z}
{s''s''_z}}}
\right.
\kern-\nulldelimiterspace} {{s''s''_z}} \right\rangle$ in the commutation
relations
(\ref{5.1}).


\section {Constructing the algebra}

In Sect. 2 the algebra $\cal G$ was defined in terms of strings of repeated
commutators of KM operators. Since all the KM operators $L_{m0}$ with
$m\geq 3$ and
$L_{0n}$ with $n\geq 4$ can be generated by repeated commutators between
operators from
the restricted set $L_{01}$, $L_{02}$, $L_{20}$, and $L_{03}$, the algebra
${\cal G}$ can
in fact be generated by strings of commutators involving only these four
operators.
It is convenient to define the subalgebra ${\cal G}'$ of the algebra ${\cal
G}$ as the
algebra generated by repeated commutators of only the three operators $L_{02}$,
$L_{20}$, and $L_{03}$. One important observation is that the commutators
of these
three operators with the operators of order $(m,n)$  only lead to operators
of of order $(m',n')$ with $m'+n'=m+n$ or $m'+n'=m+n+1$. This makes it
possible to
generate new elements of the algebra step by step in the variable $m+n$.
Thus, if the
operators at level
$m+n$ are known, all operators at level $m+n+1$ can be generated by first
commuting
these with $L_{03}$ and then (a finite number of times) with $L_{20}$ and
$L_{02}$. The
remaining difficulty is to establish the identities between operators
corresponding to
different orderings of $a_i^{\dagger}$ and $a_i$, as discussed above. The
extension of
the algebra ${\cal G}'$ to the full algebra ${\cal G}$ is done simply by
including the
generators $L_{01}$, $L_{10}$ and $L_{00}$. No new operators are generated
to higher
order $(m,n)$, due to the simple form of the commutators between $L_{01}$
and the
operators $L_{02}$, $L_{20}$, and $L_{03}$.

The commutator between a KM operator and another operator of well-defined
order can be
viewed as a translation in the lattice formed by the points $(m,n)$. A string
of
repeated operators then is represented as a path in the lattice. When
constructing the
operators of the algebra at a given point $(m,n)$, it is useful to
determine the
independent paths, constructed from the three operators
$L_{02}$, $L_{20}$, and $L_{03}$, which begin with either a KM or Virasoro
operator and
lead to the point $(m,n$).
By {\em independent paths} we then mean that the operators generated by
these paths
cannot be related by use of the Jacobi identity and/or the commutation
relations
satisfied by $L_{02}$, $L_{20}$ and $L_{03}$. This restriction to
independent paths
makes it possible to reduce the number of commutators to be evaluated when
constructing operators at the point $(m,n)$. For the set of operators
derived in this
way, the next step is to project out the spin components, and if more than
one operator
of a given spin is generated at a given site, one has to check for possible
linear
dependence between the operators.

We have illustrated the independent paths in Fig.~2 for the special case
$(m,n)=(3,3)$. A detailed check of the identities between one-particle
operators and of
the linear dependences between operators belonging to the algebra $\cal G$
is performed
in the Appendix A for the case $(m,n)=(2,4)$. Note that due to identities
between
one-particle operators, the expressions found for the operators, in terms of
$a$ and $a^\dagger$, in general will not be unique.


\begin{figure}[htb]
\unitlength=.75mm
\thicklines
\begin{center}
\begin{picture}(120,130)
\put(10,10){\vector(1,0){115}}
\put(10,10){\vector(0,1){115}}
\put(20,9){\line(0,1){2}}
\put(35,9){\line(0,1){2}}
\put(50,9){\line(0,1){2}}
\put(65,9){\line(0,1){2}}
\put(80,9){\line(0,1){2}}
\put(95,9){\line(0,1){2}}
\put(110,9){\line(0,1){2}}
\put(9,20){\line(1,0){2}}
\put(9,35){\line(1,0){2}}
\put(9,50){\line(1,0){2}}
\put(9,65){\line(1,0){2}}
\put(9,80){\line(1,0){2}}
\put(9,95){\line(1,0){2}}
\put(9,110){\line(1,0){2}}
\put(130,10){\makebox(0,0){$n$}}
\put(10,130){\makebox(0,0){$m$}}
\put(20,20){\begin{picture}(100,100)
\put(0,0){\circle{2}}
\put(0,-15){\makebox(0,0){$0$}}
\put(15,-15){\makebox(0,0){$1$}}
\put(30,-15){\makebox(0,0){$2$}}
\put(45,-15){\makebox(0,0){$3$}}
\put(60,-15){\makebox(0,0){$4$}}
\put(75,-15){\makebox(0,0){$5$}}
\put(90,-15){\makebox(0,0){$6$}}
\put(-15,0){\makebox(0,0){$0$}}
\put(-15,15){\makebox(0,0){$1$}}
\put(-15,30){\makebox(0,0){$2$}}
\put(-15,45){\makebox(0,0){$3$}}
\put(-15,60){\makebox(0,0){$4$}}
\put(-15,75){\makebox(0,0){$5$}}
\put(-15,90){\makebox(0,0){$6$}}
\put(15,0){\circle{2}}
\put(30,0){\circle{2}}
\put(45,0){\circle{2}}
\put(60,0){\circle{2}}
\put(75,0){\circle{2}}
\put(90,0){\circle{2}}
\put(0,15){\circle{2}}
\put(15,15){\circle{2}}
\put(30,15){\circle{2}}
\put(45,15){\circle{2}}
\put(60,15){\circle{2}}
\put(75,15){\circle{2}}
\put(0,30){\circle{2}}
\put(15,30){\circle{2}}
\put(30,30){\circle{2}}
\put(45,30){\circle{2}}
\put(60,30){\circle{2}}
\put(0,45){\circle{2}}
\put(15,45){\circle{2}}
\put(30,45){\circle{2}}
\put(45,45){\circle*{2}}
\put(0,60){\circle{2}}
\put(15,60){\circle{2}}
\put(30,60){\circle{2}}
\put(0,75){\circle{2}}
\put(15,75){\circle{2}}
\put(0,90){\circle{2}}
\put(3,58.5){\vector(2,-1){24}}
\put(33,43.5){\vector(2,-1){24}}
\put(18,58.5){\vector(2,-1){24}}
\put(58,32){\vector(-1,1){11}}
\put(17,73){\vector(1,-1){11}}
\put(32,58){\vector(1,-1){11}}

 \end{picture}}

\end{picture}
\end{center}
\caption{Three independent paths to the point $(3,3)$.}
\label{fig2}
\end{figure}
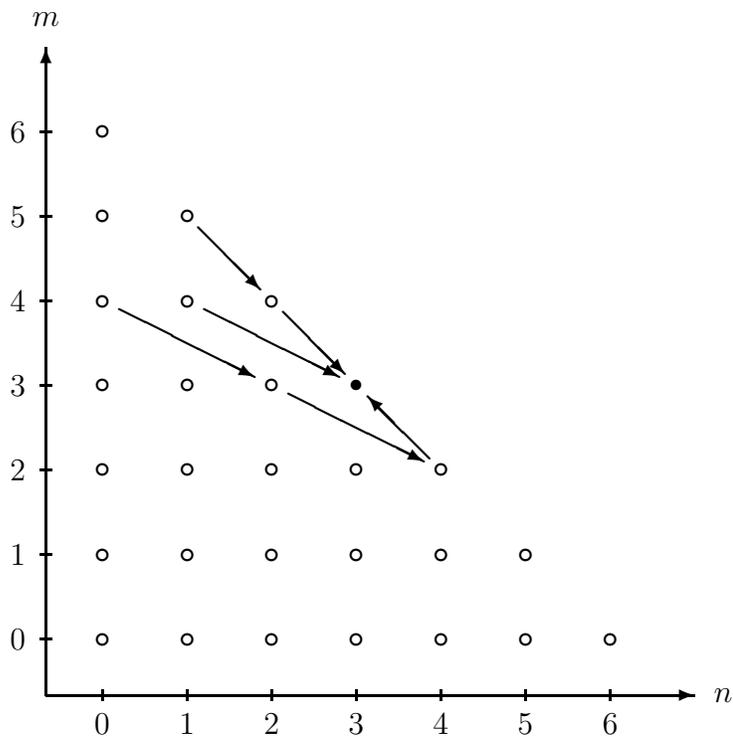


We now examine the operators of the algebra $\cal G$ for the special cases
$m+n\leq 6$. If
$m=0$ or $n=0$ there is only one operator (the Kac Moody operator). Also at
$(m,1)$ and
$(1,n)$ there is only one mirror-symmetric operator, the Virasoro operator
(\ref 12).
This is readily shown by use of (\ref {15}). We first note that, since
there is only
one (trivial) operator for $m+n=0$, there is a single spin $0$ operator of
this order.
This operator commutes with all operators of the algebra, and we interpret
it as the
particle number operator. To $m+n=1$ there corresponds a single spin $1/2$
multiplet.
The operators are
\be
L_{10}&=&a^{\dagger}\,, \nonumber \\
L_{01}&=&a \,.
\ee Here we have used a
notation where
the summation over particle indices is suppressed. In the following we will
use this
convention. This should make no confusion, since we only refer to one-particle
operators. At level $m+n=2$ we have the spin $1$ multiplet which we have
used to define the spin content of the algebra,
\be
L_{20}&=&a^{\dagger 2}\,,	\nonumber \\
L_{11}&=& \half (a^\dagger a + a a^\dagger)\,, \nonumber \\
L_{02}&=&a^{ 2}\,.
\ee
Also the operators at $m+n=3$ are known, since only KM and Virasoro
operators are
included in the corresponding multiplet. This multiplet is unique and has
spin $3/2$,
\be
L_{30}&=&a^{\dagger 3}\,, \nonumber \\
L_{21}&=& \half (a^{\dagger 2} a + aa^{\dagger 2})\,.
\ee
Hereafter we only write the components of multiplets with $s_z\geq 0$. The
operators with $s_z\leq 0$ are obtained (for $m+n\leq 6$) by the Hermitian
conjugation:
\be
L^{s}_{nm}=(L^{s}_{mn})^{\dagger}\,.
\ee

\noindent {\em The case $m+n=4$}:

In this case we need to evaluate the operators of
order $(2,2)$. There are two ways to generate operators of this order, by
taking the
commutators $[L_{02},L_{31}]$ and $[L_{03},L_{30}]$. The first one
generates a spin 2 operator and the second one a linear combination of a
spin $2$ and a
spin $0$ operator. We find the following expression for the spin 2 multiplet
\be
L_{40}&=&a^{\dagger 4}\,,	\nonumber \\
L_{31}&=& \half (a^{\dagger 3} a + ms) \,,\nonumber \\
L^{2}_{22}&=&\frac{1}{6} (a^{\dagger 2}a^2 +a^{\dagger} a
a^{\dagger} a+a^\dagger a^2 a^\dagger+ms)\,.
\ee
In these expressions a simplified notation is
introduced, where $ms$ denotes the mirror-symmetric operator. The spin $0$
operator has
the form
\be
L^{0}_{22}=a^{\dagger 2}a^2 -a^{\dagger} a a^{\dagger} a + ms \,.
\ee

\noindent {\em The case $m+n=5$}:

In this case we have to calculate the operators of orders (3,2) or (2,3).
Consider
order (3,2). We generate the operators of this order by taking the commutators
$[L_{02},L_{41}]$ and $[L_{03},L_{40}]$. The first one gives a spin $5/2$
operator and
the second one a linear combination of a spin $5/2$ and a spin $1/2$
operator. For the
spin $5/2$ multiplet we have the following expressions for the operators,
\be
L_{50}&=&a^{\dagger 5}	\,,\nonumber \\
L_{41}&=&\half (a^{\dagger 4} a + ms)\,, \nonumber \\
L^{5/2}_{32}&=&\frac{1}{4}(a^{\dagger 3}a^2
+2 a^{\dagger} a a^{\dagger 2} a- a^{\dagger 2} a a^\dagger a + ms)\,,
\ee
and after separating out the spin $1/2$ components, we find for this
multiplet the operators
\be
L^{1/2}_{32}&=&\frac12(a^{\dagger 3}a^2 - a^{\dagger 2} a a^\dagger a +
ms)\,.
\ee

\noindent {\em The case $m+n=6$}:

We examine the independent operators of order $(3,3)$. All other operators
can be
derived from these by use of the lowering or raising operators. We have to
consider
three different commutator expressions, $[L_{02},[L_{02},L_{51}]]$,
$[L_{03},L_{41}]$
and $[L_{20},[L_{03},[L_{03},L_{51}]]]$. The first one defines a spin $3$
and the two
others linear combinations of spin $3$ and spin $1$ operators. After
projecting out the
different spin components and checking for possible linear dependence (see
Appendix A), we find that there are two independent spin $1$ multiplets in
addition to
the spin $3$ multiplet. The explicit expressions for the spin $3$ operators
are
\be
L_{60}&=&a^{\dagger 6}\,,\nonumber \\
L_{51}&=&\half (a^{\dagger 5} a + ms) \,,\nonumber \\
L^{3}_{42}&=&\frac{1}{10}(a^{\dagger 4}a^2+a^{\dagger 2}a a^{\dagger 2} a+
2a^{\dagger}a a^{\dagger 3}a+a^{\dagger 3}a^2a^{\dagger} +ms) \,,\nonumber\\
L^{3}_{33}&=&\frac{1}{20}(3 a^{\dagger 3} a^3- 3a^{\dagger 2} a a^\dagger
a^2 + 2
a^{\dagger} a a^{\dagger 2} a^2 + 2a^{\dagger 2} a^2 a^\dagger a + a^\dagger a
a^\dagger a a^\dagger a +5a^\dagger a^2 a^{\dagger 2} a + ms). \nonumber \\
\ee
For the first spin $1$
multiplet, we find
\be L^{1a}_{42}&=&\frac{1}{6}(5a^{\dagger
4}a^2-7a^{\dagger 3}a
a^{\dagger} a +3a^{\dagger 2}a a^{\dagger 2} a+a^{\dagger}a a^{\dagger
3}a-2a^{\dagger
3}a^2a^{\dagger}+ms)\,,\nonumber\\
L^{1a}_{33}&=&\frac{1}{4}( a^{\dagger} a a^{\dagger 2} a^2 +
a^{\dagger 2}
a^2 a^\dagger a -2 a^\dagger a a^\dagger a a^\dagger a +ms)\,,
\label{a}\ee
and for the
second multiplet
\be
L^{1b}_{42}&=&\frac{1}{6}(3a^{\dagger 4}a^2-5a^{\dagger 3}a a^{\dagger} a
+3a^{\dagger 2}a a^{\dagger 2} a+a^{\dagger}a a^{\dagger 3}a-2a^{\dagger
3}a^2a^{\dagger}+ms) \,,\nonumber\\
L^{1b}_{33}&=&-\frac{1}{6}( a^{\dagger 3} a^3-
a^{\dagger 2} a a^\dagger a^2 - a^{\dagger} a a^{\dagger 2} a^2 -a^{\dagger
2} a^2a^\dagger a + 2 a^\dagger a a^\dagger a a^\dagger a +ms)\,.
\label{b}\ee
The distinction
between the two spin $1$ multiplets is not unique. The $b$ multiplet has
been chosen to
vanish for $\lambda=0$, while the $a$ multiplet gets a simple form in the
subspace
where $K_{ij}=1$, as will be shown below.

In Appendix B we summarize some of the commutation relations which involve the
low-order operators. We stress the point that the commutation relations
between the
operators $L_{mn}$, which are derived from the explicit form of these in the
Calogero model, by definition characterize the (abstract) algebra $\cal G$. The
operators $L_{mn}$ of the Calogero model, in turn,  can be viewed as defining a
representation of this algebra. Other representations may exist which are
also of
interest from the point of view of generalized statistics. However, for such
representations the expressions for the operators $L_{mn}$ given above may not
be valid.

The operators $L_{mn}$ of the extended Calogero model define a {\em reducible}
representation of $\cal G$. This follows from the fact that all operators
commute with
the permutations $K_{ij}$. Restriction of the model to a subspace which
defines an
irreducible representation of $\cal G$ then will correspond to a
restriction of the
operators $K_{ij}$ to an irreducible representation of the permutation
group. We note
that such a restriction is needed when we consider the model as describing
a system of
{\em identical} particles.

When the operators $L_{mn}$ are restricted to a subspace which defines an
irreducible
representation of $\cal G$, there will be more identities between the
operators than
those which are satisfied in the general case. We will exemplify this for
the low-order
operators in the irreducible representations characterized by $K_{ij}=1$,
{\em i.e.} for
representations which are symmetric in the particle indices.

As already noticed, the spin $0$ operator $L_{00}$ can be identified with
the particle
number $N$. We find that also the second spin $0$ operator in this irreducible
representation is proportional to the identity operator, and can be written
as a function
of
$N$ and $\lambda$,
\be
L_{22}^0=\left(1-\lambda(\lambda-1)(N-1)\right)N \,.
\label{L22}\ee
The operator $L_{22}^0$, in the same way as $L_{00}$, commutes with all
elements of
the algebra $\cal G$. There are relations of the same form as (\ref{L22})
also for the
spin $1/2$ and spin $1$ multiplets. We have
\be
L_{32}^{1/2}=\left(1-\lambda(\lambda-1)(N-1)\right)L_{21}
\ee
and
\be L_{42}^{1a}=\left(1-\lambda(\lambda-1)(N-1)\right)L_{20}\,,
\ee
with similar
expressions for the other operators of the two spin multiplets. This
suggests that a
relation of this form may be valid also for multiplets of higher spins. For
the second spin $1$ multiplet we find
\be
L_{42}^{1b}=-\frac{1}{3}\lambda(\lambda-1)(N L_{20}-(L_{10})^2)\,.
\ee
We note here in particular the non-linear dependence of the spin
$1/2$ operator $L_{10}$.


\section{The Bose/Fermi algebra ${\cal G}_0$}
We now consider the special case
$\lambda=0$. This case describes bosons if the state space is restricted to
states
which are symmetric under interchange of particle indices and fermions if it is
restricted to antisymmetric states. We make a change in notation for this
special case,
$a_i\to b_i$, $a_i^{\dagger}\to b_i^{\dagger}$ and $L_{mn}\to \ell_{mn}$,
and the new
operators $b_i$ and $b_i^\dagger$ then satisfy the standard commutation
relations
\be
[b_i,b_j^{\dagger}]=\delta _{ij}\,.
\ee
These commutation relations imply more (linear) identities for the one-particle
operators than those satisfied for general $\lambda$. In fact, a general
mirror-symmetric one-particle operator of order $(m,n)$ now can be expanded
in the
following way
\be
\ell_{mn} = \sum_{k=0}^{k_{\rm max}} c_k(b^{\dagger m-2k}b^{n-2k}+ms)\,,
\label{lmn}\ee
with
\be k_{\rm max}(m,n)=\left[ \frac{{\rm min}(m,n)}2 \right] \,,
\label{4.2} \ee
where the
square brackets indicate the integer part. This gives the following number of
independent operators of order $(m,n)$,
\be
g(m,n)=k_{\rm max}(m,n)+1 \,.
\ee
These degeneracies determine the spin multiplets which are present for a
given value of $m+n$. The spin values are
\be
s=s_{\rm max}(m,n)-2k \,,
\ee
with
$s_{\rm max}(m,n)$ given by (\ref{smax}) and $k=0,1,\dots,k_{\rm
max}(m,n)$, with only
one operator for each spin. Hence for $\lambda=0$ there is no degeneracy
except the one
referring to different spins, and we may drop the index $\alpha$ of the
operators,
$\ell_{mn}^{s\alpha }\to \ell_{mn}^{s}$.

The operators $\ell_{mn}^{s}$ define an algebra ${\cal G}_0$, which is
smaller than the
algebra
$\cal G$ for general $\lambda$. We refer to this as the Bose/Fermi algebra.
It is
obtained from $\cal G$ by introducing the additional identities, for
operators  of
{\em the same order} $(m,n)$ and the same value of the spin $s$, which can
be derived
from the Heisenberg algebra satisfied by $b_i$ and $b_i^\dagger$. Thus, for
arbitrary
$\lambda$, there are generally several multiplets of the same spin  for
fixed $(m,n)$.
One can choose these multiplets in such a way that all of these except one
vanish for
$\lambda=0$, when the additional identities are introduced. We label the
non-vanishing
multiplet by letter $a$. The identities then can be written as
\be
\ell_{mn}^{sa}=\ell_{mn}^{s}\;;\;\;\;\;
\ell_{mn}^\alpha=0\; \;\;\;{\rm for}\;\;\; \alpha\neq a  \,.
\label{reduction-a}
\ee
For general $\lambda$, the operators that vanish for $\lambda=0$ form an
invariant
subalgebra of $\cal G$. This follows from the fact that the operators
$\ell_{mn}^s$ satisfy the commutation relations of $\cal G$ (since this is
$\lambda$-independent) as well as the commutation relations of ${\cal
G}_0$. The algebra
${\cal G}_0$ can be  obtained from $\cal G$ by division with this invariant
subalgebra.
This implies that there is a connection between the structure constants of
the two algebras.

The commutation relations of the Bose/Fermi algebra ${\cal G}_0$ have the form,
\be
\left[\ell^{s}_{mn},\ell^{s'}_{m'n'}\right]=\sum_{s''=|s-s'|}^{s+s'}
c_{ss_z,s's'_z}^{s''}\ell^{s''}_{m+m'-1,n+n'-1}\,.
\label{5.1b}\ee
where $c_{ss_z,s's'_z}^{s''}$ is vanish unless $s''-s-s'$ is a positive odd
integer.
The structure constants can be derived from those of the algebra
$W_{1+\infty}$, as we will discuss below. They in turn determine some of
the structure
constants (\ref{5.1}) of ${\cal G}$, namely those referring only to the
$a$-multiplets.
We have
\be
d_{s\,s_{\rm max}\,a,s'\,s'_{\rm max}\,a}^{s''a}
\left\langle{{ss's_zs'_z}}
\mathrel{\left | {\vphantom {{ss's_zs'_z} {s''s''_z}}} \right.
\kern-\nulldelimiterspace} {{s''s''_z}}
\right\rangle
=c_{ss_z,s's'_z}^{s''} \,.
\label{}
\ee
Thus, some of the structure constants of the full algebra $\cal G$ can be
determined by
calculations of the commutators in the simpler case $\lambda=0$.

The structure constants $c_{ss_z,s's'_z}^{s''}$ do not depend on the
variable $m+n$. In
fact they are directly related to the structure constants of the smaller
algebra
$W_{1+\infty}$. To see this we note that there are more identities for
$\lambda=0$ than
those valid for fixed $(m,n)$. The expansion (\ref{lmn}) shows that all
except one
operator of order $(m,n)$ can be expressed as a linear combination of
operators of
lower orders. The operator which cannot be expressed in terms of operators
of lower
orders clearly must be the one with maximal spin, and operators with other
spin values
are identified with operators of the same spin but with lower values of
$m+n$. We write
these identifications as
\be
\ell^{s,s_z}=\ell_{mn}^s\;, \;\;\;\; s_z=\half(m-n) \,.
\label{4.2a}
\ee Thus, when all the linear relations between operators for $\lambda=0$
are taken
into account there is only one independent operator associated with each
order $(m,n)$,
and there is only one independent spin multiplet for each integer or half
integer value
in the algebra. This algebra is the one which is referred to as the
$W_{1+\infty}$
algebra of the Bose/Fermi system. The commutation relations of this algebra
can be
written as
\be
\left[\ell^{s,s_z},\ell^{s',s'_z}\right]=\sum_{s''=|s-s'|}^{s+s'}
c_{ss_z,s's'_z}^{s''}\ell^{s'',s_z+s'_z}\,.
\label{5.4} \ee
with $s'+s-s''$ odd positive,
and the identification (\ref{4.2a}) then gives the corresponding expression
(\ref{5.1b}) for the larger algebra ${\cal G}_0$. Even if the two sets of
commutation
relations (\ref{5.1b}) and (\ref{5.4}) are closely related, it is of
interest to note an
important difference. If we introduce Planck's constant $\hbar$, the RHS of
(\ref{5.4})
can be viewed as an expansion in powers of $\hbar$. On the other hand, the
commutation
relations of ${\cal G}_0$, in the same way as for $\cal G$, do not contain
higher orders
in
$\hbar$.

For the operators $\ell^{s,s_z}$ we can find a general expression in terms
of $b$ and
$b^\dagger$. To find this we make use of the commutator (\ref{30}) which
now can be
written as,
\be
[\ell^{1,-1},\ell^{s,s_z}]=2(s+s_z)\ell^{s,s_z-1}\,.
\ee
Starting from the operator $\ell^{s,s}$, one can use this recursively to
derive the
following expression for the operator $\ell^{s,s_z}$,
\be
\ell^{s,s_z}=\sum_{m=0}^{s-s_z}
\frac{(s+s_z)!(s-s_z)!}{2^m m!(s+s_z-m)!(s-s_z-m)!}
(b^{\dagger})^{s+s_z-m}b^{s-s_z-m} \,.
\label{5.2b}\ee
This expression can be used to derive the structure constants in
(\ref{5.4}) for the algebra $W_{1+\infty}$ in the basis of the operators
$\ell^{s,s_z}$. We quote here only the first term (with highest spin)  in
the RHS of (\ref{5.4}),
\be
[\ell^{s,s_z},\ell^{s',s'_z}]=2(ss'_z-s's_z)\ell^{s+s'-1,s_z+s'_z}+ \cdots
\label{ss'1}
\ee

The  operators $\ell^{s,s_z}$ define a representation of the algebra ${\cal
G}_0$. This follows from the fact that the identifications (\ref{4.2a}),
$\ell^{s}_{m,n}=\ell^{s,s_z}$, are consistent with the commutation
relations of the
algebra. It is of interest to note that another set of identifications,
$\ell^{s}_{m,n}=\ell^{s,s_z}\delta_{m,s+s_z}\delta_{n,s-s_z}$, is also
consistent with
the commutation relations of ${\cal G}_0$.  These identifications
correspond to keeping
only the operator of the maximal spin at each point
$(m,n)$. In this case all terms in (\ref{ss'1}) except the first one
vanish, and with
$s$ and $s_z$ related to $m$ and $n$ as in the maximal spin multiplet, the
commutation
relations are reduced to that of the classical algebra $w_\infty$,
\be
[\ell_{mn},\ell_{m'n'}]=(nm'-n'm)\ell_{m+m'-1,n+n'-1}\,.
\ee
Thus, representations of the algebras $W_{1+\infty}$ and $w_\infty$
also provide representations of the larger algebras ${\cal G}_0$ and $\cal G$.

\section{Conclusions}

To summarize, we have shown that a parameter-independent algebra ${\cal G}$ of
one-particle operators can be defined for the Calogero model. A basis for
this algebra
is generated by repeated commutators between the set of operators $L_{01}$,
$L_{02}$,
$L_{03}$, and $L_{20}$. We have given a recursive way to construct the
algebra and
used it to explicitly construct the generators for low orders. The elements
of the
algebra are represented in terms of spin operators, $L_{mn}^{s \alpha}$,
and are grouped
into spin multiplets. We have shown that in the Bose and Fermi cases one
can define a
smaller algebra ${\cal G}_0$ of a similar form as ${\cal G}$. This algebra
is closely
related to the algebra
$W_{1+\infty}$, but also to the algebra $w_\infty$ of the classical system.

We interpret the algebra ${\cal G}$ as the algebra of observables for a
system of
identical particles which satisfy generalized statistics in one dimension.
The algebra
is independent of particle number and of the statistics parameter. A given
system of
particles corresponds to an irreducible representation of the algebra. Such a
representation is characterized by the particle number and by a fixed value
of the
statistics parameter. Identities exist between the operators in such an
irreducible
representation, and these identities reduce the number of independent
observables. An
interesting question is whether this reduction is sufficient to match the
number of
degrees of freedom of the (classical) system. The expressions found for
low-order
operators in the representations $K_{ij}=1$ may indicate that this is in fact
the case.

Since the structure of the algebra $\cal G$ is only partly known, it will be of
interest to investigate this structure further. The algebra clearly can be
constructed
step by step in the way discussed in this paper. However the aim would be
to try to
find more general expressions for the algebraic relations. Let us finally
mention again
the question of other representations of the algebra $\cal G$, different
from those
found in the Calogero model. There may exist such representations which are
also
interesting from the point of view of generalized statistics in one dimension.

\bigskip
\centerline{{\bf Acknowledgments}}

\smallskip
We are grateful to U. Lindstr{\"o}m, J. Myrheim, A.P. Polychronakos, R.
Varnhagen, and M.
Vasiliev for useful discussions. We would like to express special thanks to
Jan Myrheim
whose computer calculations helped a lot in clarifying the structure of the
discussed
algebra.

S.B.I. gratefully acknowledges warm hospitality of Department of Physics of
University
of Oslo. The work of S.B.I. was in part supported by the Russian Foundation for
Fundamental Research under Grant No. 95-02-04337.


\appendix
\section{\bf Appendix}

Here we outline the way of finding all the linearly independent elements of the
algebra $\cal G$ for the order $(2,4)$. These operators can be generated in
the following
ways:

\begin{itemize}
\item[(i)] starting from $L_{51}$, by repeated commutations with $L_{02}$, or
equivalently, by taking the commutator of $L_{20}$ with $L_{15}$;

\item[(ii)] ``transporting'' the two linearly independent operators from
the point
(3,2) by taking the commutators of $L_{03}$ with each of those operators.
\end{itemize}
This yields the three operators
\be
L^{(1)}&=&A+B+D+F+I \,,\nonumber\\
L^{(2)}&=&B+C+F\,,\nonumber\\
L^{(3)}&=&2A+2D+G+I\,,
\label{operL}\ee
where
\be
2A&=&a^4a^{{\dagger}2} +ms\,,\nonumber\\
2B&=&a^3a^{{\dagger}}aa^{\dagger}+ms\,,\nonumber\\
2C&=&a^3a^{{\dagger}2}a +ms\,,\nonumber\\
2D&=&a^2a^{{\dagger}}a^2a^{{\dagger}}+ms\,,\nonumber\\
2E&=&a^2a^{{\dagger}}aa^{\dagger}a+ms\,,\nonumber\\
2F&=&aa^{\dagger}a^3a^{{\dagger}} +ms\,,\nonumber\\
G&=&a^2a^{{\dagger}2}a^2 \,,\nonumber\\
H&=&aa^{\dagger}a^2a^{{\dagger}}a \,,\nonumber\\
I&=&a^{{\dagger}}a^4a^{\dagger} \label{oper}\ee
are mirror-symmetric operators of order (2,4). The set (\ref{oper})
includes all the mirror-symmetric operators of this order.

There are four identities between the operators (\ref{oper})
\be
B+I&=&C+F\,,\nonumber\\
A+F+I&=&C+E+G\,,\nonumber\\
A+3E&=&3B+I\,,\nonumber\\
D+F&=&E+H\,,
\label{iden}\ee
which can be
derived using the identity (\ref{15a}) as well as using the Jacobi identity to
rearrange expressions with repeated commutators. Using these four identities
(\ref{iden}), one can express the three operators (\ref{operL}) in
terms of only five of the operators (\ref{oper}), {\em e.g.} $B$, $C$, $D$,
$E$, and
$F$. If this is a linearly independent set of operators also the  operators
$L^{(1)}$,$L^{(2)}$, and $L^{(3)}$ will be linearly independent. However,
we will make
an independent check of whether there exists any additional identity, not
included among
the four identities (\ref{iden}), which would make the three operators
linearly dependent.

In the Bose/Fermi case $\lambda =0$ one can easily show that the three
operators obtained
in this way are linearly dependent.  This is due to an additional identity of
the form
\be B+C+3D-9E+4F=0\,.
\label{iden-bos}\ee
If the operators (\ref{operL}) are linearly dependent in the
general case, this identity has to be valid for general $\lambda$.

To check whether the identity (\ref{iden-bos}) is true or not in the
general case, we
consider the action of the operators (\ref{oper}) in the space of symmetric
functions. We
first extend the definition of mirror symmetry to any product of the
operators $a_i$,
$a_j^{\dagger}$ and $K_{ij}$: the mirror-symmetric expression is
obtained by inversion of the order of the operators in the product. We
rewrite the
operators (\ref{oper}) in a {\em normal-ordered mirror-symmetric} form. For
that ---
consider {\em e.g.} the operator $B=\frac12 (a^{{\dagger}}aa^{\dagger}a^3 +
ms)$ --- we rewrite the first term in a normal-ordered form and the second,
mirror-symmetric term, in an anti-normal-ordered form. Next, consider the
following
projection (which will be denoted by $\pi_+$): we move all the operators
$K_{ij}$ to the
right of the expression and then replace $K_{ij}$ by one. As a result, the
projection of
any of the operators (\ref{oper}) takes the form
\be
\frac12 (a^{{\dagger}2}a^4+ms) + \alpha a^{2} + \lambda\left[\beta a^{2} +
\gamma \mathop{{\sum}'}_{il} a_ia_l\right]\,,
\label{proj}\ee where $\alpha$ is
independent of
$\lambda$, and $\beta$ and $\gamma$ are linear functions of $\lambda$; the
prime means
summation only over non-equal $i$ and $l$.

For our purposes it is sufficient to make a restriction to first order in
$\lambda$ in (\ref{proj}). Introducing the basis
\be
e_0=\frac12 (a^{{\dagger}2}a^4+ms),\;\;
e_1=\frac12 a^{2},\;\;
e_2=\frac12 \mathop{{\sum}'}_{il}(a_i^2+a_ia_l+a_l^2)\,,
\ee
 one obtains
\be
\pi_+(A)&=&e_0\,,\nonumber\\
\pi_+(B)&=&e_0-3e_1-\lambda e_2\,,\nonumber\\
\pi_+(C)&=&e_0-6e_1-2\lambda e_2-3\lambda (N-1)e_1\,,\nonumber\\
\pi_+(D)&=&e_0-6e_1-2\lambda e_2-\lambda (N-1)e_1\,,\nonumber\\
\pi_+(E)&=&e_0-7e_1-3\lambda e_2\,,\nonumber\\
\pi_+(F)&=&e_0-9e_1-5\lambda e_2+3\lambda (N-1)e_1\,,\nonumber\\
\pi_+(G)&=&e_0-8e_1-6\lambda e_2+6\lambda (N-1)e_1\,,\nonumber\\
\pi_+(H)&=&e_0-8e_1-4\lambda e_2+2\lambda (N-1)e_1\,,\nonumber\\
\pi_+(I)&=&e_0-12e_1-6\lambda e_2\,.
\ee
With these expressions, the identities
(\ref{iden}) projected onto the space of symmetric functions are satisfied
as it should be. On the other hand, we get
\be
\pi_+(B+C+3D-9E+4F)=10\lambda e_2 +6\lambda (N-1)e_1\,,
\ee
which shows that the identity (\ref{iden-bos}) holds only for $\lambda=0$.
This proves
that the three operators $L^{(1)}$, $L^{(2)}$, and $L^{(3)}$ for general
$\lambda$ are
linearly independent, {\em i.e.} the degeneracy at point (2,4) is
$g(2,4)=3$. Due to the
obvious symmetry between the points $(m,n)$ and $(n,m)$, we also have
$g(4,2)=3$.

Similar considerations show that $g(3,3)=3$. Note that three linearly
independent
operators of order (3,3) can be obtained by taking the commutators of
$L_{20}$ with the
three linearly independent operators at point (2,4).

\section{Appendix}

Here we  discuss the commutation relations for the the spin operators
 introduced in Sec.~4. We start with the commutators between the operator
$L_{03}$ and operators of maximal spin. Consider first the commutator of
the form $[L_{03},L_{m0}]$ for $1\leq m\leq 5$.
It is seen from (\ref{10}) that this commutator is an operator
of order $(m-1,2)$. The calculation of the spin multiplets for operators
of orders $(m,n)$ with $m+n\leq 6$ imply that operators of order $(m-1,2)$ are
linear combinations of operators of spins $\frac12(m+1)$ and $\frac12(m-3)$.
For $1\leq m\leq 4$, there is only one multiplet of spin $\frac12(m-3)$. We
find explicitely for this case
\be
[L_{03},L_{m0}]=3mL_{m-1,2}^{(m+1)/2}+
\frac14 m(m-1)(m-2)L_{m-1,2}^{(m-3)/2}\,.
\label{L03Lm0}\ee
For $ m=5 $, there are two independent operators of spin
$\frac12(m-3)$ of order $(m-1,2)$. In this case, instead of the second term in
the RHS of (\ref{L03Lm0}), we obtain a linear combination of these two
operators:
\be [L_{03},L_{50}]=15L_{42}^3+15L_{42}^{1a}-18L_{42}^{1b}\,.
\label{L03L50}\ee

In a similar way, we find
\be [L_{03},L_{41}]=12L_{33}^{3}+
6L_{33}^{1a}-\frac{36}{5}L_{33}^{1b} \label{L0341}
\ee
and
\be
[L_{03},L_{32}^{5/2}]=9L_{24}^{3}+\frac32L_{24}^{1a}-\frac95 L_{24}^{1b}\,.
 \label{L0332}\ee

Next, consider commutators between the operator $L_{03}$ and operators of a
non-maximal spin. We first observe that
\be
[L_{03},L_{22}^0]=0\,.
\label{L03L220}\ee In fact, the operator $L_{22}^0$ commutes with every
operator of the algebra $\cal G$. This follows from the representation of
the operator  $L_{22}^0$ in the form
\be
L_{22}^0= N-\lambda^2N(N-1)+\lambda\sum_{il}K_{il}
\ee
and the observation that the operator $K_{il}$ commutes with any
one-particle operator.

We also have
\be [L_{03},L_{32}^{1/2}]=3L_{24}^{1a}
\label{L03L3212}\ee
and
\be [L_{03},L_{23}^{1/2}]=0\,.
\label{L03L2312}\ee
The last equality reflects the selection rule implied by
the spin addition formula.

Finally, we consider  commutators involving the operators $L_{01}$ and
$L_{10}$.
For all the operators $L_{mn}^s$, with $m+n\leq 6$, that do not vanish for
$\lambda=0$, we obtain
\be [L_{01},L_{mn}^s]=\frac12(2s+m-n) L_{m-1,n}^{s-1/2}\,.
\label{L01}\ee
On the other hand, for the operators of the multiplet (\ref{b}) at level
$m+n=6$, which
vanish for $\lambda=0$, we have
\be
[L_{01},L_{mn}^{1b}]=[L_{10},L_{mn}^{1b}]=0\,.
\label{L10b}\ee
Note that the last equality does not follows from the selection rule
dictated by the spin
addition formula. It is rather a conseqence of the fact that the operators
which
vanish for $\lambda=0$ form an invarant subalgebra. The operators
$L_{mn}^{1b}$ belong to
this subalgebra, and the commutator then also is an element of the
subalgebra. Since
the commutator is an operator of level $m+n=5$ and there are no elements of the
invariant subalgebra at this level except the trivial one, the commutator
has to vanish.

For low orders we also find the additional rule that the commutators  between
operators of spin
$s$ and
$s'$ only include operators of spins
$s+s'-1-2k$ with $k$ non-negative integer ({\em i.e.} operators of spins
differing in
steps of 2). The vanishing commutator (\ref{L03L220}) is a special case of
this rule.
The additional selection rules that do not follow from the spin selection rules
suggest the presence of more symmetry than we have established explicitly.



\newpage
\begin{thebibliography}{99}

\bibitem{LM-Heis} J. M. Leinaas and J. Myrheim, Phys. Rev. B {\bf 37}
(1988) 9286; Int.
J. Mod. Phys. B {\bf 5} (1991) 2573; Int. J. Mod. Phys. A {\bf 8} (1993) 3649.
\bibitem{V}  M. Vasiliev, Int. J. Mod. Phys. A {\bf 6} (1991) 1115.
\bibitem{Poly-PRL92} A. P. Polychronakos, Phys. Rev. Lett. {\bf 69} (1992) 703.
\bibitem{BHV} L. Brink, T. H. Hansson, and M. Vasiliev,
Phys. Lett. {\bf B 286} (1992) 109;\\
L. Brink, T. H. Hansson, S. E. Konstein, and M.
Vasiliev, Nucl. Phys. {\bf B384} (1993) 591.
\bibitem{Perelomov} A. M. Perelomov, Teor. Mat. Fiz. {\bf 6} (1971) 364.
\bibitem{PolyNPB} A. P. Polychronakos, Nucl. Phys. {\bf B324} (1989) 597.
\bibitem{H} F. D. M. Haldane, Phys. Rev. Lett. {\bf 67}, 937 (1991).
\bibitem{StatMech} S. B. Isakov, Int. J. Mod. Phys. A {\bf 9} (1994) 2563;
Mod. Phys.Lett. {\bf B 8} (1994) 319;\\
D. Bernard and Y.-S. Wu, Preprint SPhT-94-043, UU-HEP/94-03,
cond-mat/9404025;\\
Z. N. C. Ha, Phys. Rev. Lett. {\bf 73} (1994) 1574; Nucl.
Phys. {\bf 435} {\bf [FS]} (1995) 604;\\
M. V. N. Murthy and R. Shankar, Phys. Rev. Lett.
{\bf 73} (1994) 3331;\\
A. Dasni\`eres de Veigy and S. Ouvry, Mod. Phys. Lett. B {\bf 9} (1995) 271.
 \bibitem{CTZ} A. Cappelli, C. A. Trugenberger,
and G. Zemba, Nucl. Phys. {\bf B396} (1993) 465.
\bibitem{Cal} F. Calogero, J. Math. Phys. {\bf 10} (1969) 2191, 2197; {\bf 12}
(1971) 419.
\bibitem{BV} E. Bergshoeff and M. Vasiliev, Int. J. Mod. Phys. A {\bf 10}
(1995) 3477.
\bibitem{Poly-pc} A. P. Polychronakos, private communication.
\end{thebibliography}
\end{document}